\documentclass[journal]{IEEEtran}

\usepackage[pdftex]{graphicx}
\graphicspath{{../pdf/}{../jpeg/}{../png/}}
\DeclareGraphicsExtensions{.pdf,.jpeg,.png}

\usepackage{amsmath}
\usepackage{array}
\usepackage{standalone}
\usepackage{etoolbox}
\newtoggle{arXiv}
\newtoggle{bibrun}
\usepackage{newtxtext, newtxmath}
\usepackage{graphicx}
\usepackage{color, calc}
\usepackage{booktabs}
\newcolumntype{L}{>{$}l<{$}}
\newcolumntype{C}{>{$}c<{$}}
\newcolumntype{R}{>{$}r<{$}}
\usepackage{tikz, tikz-3dplot}
\usetikzlibrary{arrows, arrows.meta, calc, positioning}
\usetikzlibrary{decorations.pathmorphing}	
\tikzset{>=Stealth}
\usepackage{pgfplots}
\usepackage{siunitx, physics}
\newcommand{\Exp}[1]{\mathrm{e}^{#1}}
\usepackage{enumitem}
\setlist[description]{labelindent=0pt, leftmargin=\parindent, font=\normalfont\itshape}
\usepackage{url}
\usepackage{subfigure}
\usepackage{textcomp}
\usepackage{eurosym}
\usepackage{stfloats}
\usepackage{bm,upgreek}

\begin{document}

\title{Prepolarized MRI of Hard Tissues and Solid-State Matter}

\author{\IEEEauthorblockN{
		Jos\'e~M.~Gonz\'alez\IEEEauthorrefmark{1}$^,$\IEEEauthorrefmark{2},
		Jose~Borreguero\IEEEauthorrefmark{1}\IEEEauthorrefmark{2},
		Eduardo~Pall\'as\IEEEauthorrefmark{3},
		Juan~P.~Rigla\IEEEauthorrefmark{2},
		Jos\'e~M.~Algar\'{\i}n\IEEEauthorrefmark{3},
		Rub\'en~Bosch\IEEEauthorrefmark{2},
		Fernando~Galve\IEEEauthorrefmark{3},
		Daniel~Grau-Ruiz\IEEEauthorrefmark{2},
		Rub\'en~Pellicer\IEEEauthorrefmark{3},
		Alfonso~R\'ios\IEEEauthorrefmark{2},
		Jos\'e~M.~Benlloch\IEEEauthorrefmark{3}, and
		Joseba~Alonso\IEEEauthorrefmark{3}}
	
	\IEEEauthorblockA{\IEEEauthorrefmark{1}\emph{Equally contributing authors}}\\
	\IEEEauthorblockA{\IEEEauthorrefmark{2}Tesoro Imaging S.L., 46022 Valencia, Spain}\\
	\IEEEauthorblockA{\IEEEauthorrefmark{3}MRILab, Institute for Molecular Imaging and Instrumentation (i3M), Spanish National Research Council (CSIC) and Universitat Polit\`ecnica de Val\`encia (UPV), 46022 Valencia, Spain}\\
	
	Instituto de Instrumentación para Imagen Molecular, Centro Mixto CSIC—Universitat Politècnica de València, 46022 Valencia, Spain

\thanks{JMG and JB have contributed equally to this work.}
\thanks{Corresponding author: J. Alonso (joseba.alonso@i3m.upv.es).}}

\maketitle

\IEEEtitleabstractindextext{%
\begin{abstract}
Prepolarized Magnetic Resonance Imaging (PMRI) is a long-established technique conceived to counteract the loss in signal-to-noise ratio (SNR) inherent to low-field MRI systems. When it comes to hard biological tissues and solid-state matter, PMRI is severely restricted by their ultra-short characteristic relaxation times. Here we demonstrate that efficient hard tissue prepolarization is within reach with a special-purpose 0.26\,T scanner designed for dental MRI and equipped with suitable high-power electronics. We have characterized the performance of a 0.5\,T prepolarizer module which can be switched on and off in just 200\,$\bm{\upmu}$s. To that end, we have used resin, dental and bone samples, all with $\mathbf{T_1}$ times in the order of 20\,ms at our field strength. The measured SNR enhancement is in good agreement with a simple theoretical model, and small deviations in extreme regimes can be attributed to mechanical vibrations due to the magnetic interaction between the prepolarization and main magnets. Finally, we argue that these results can be applied to clinical dental imaging, opening the door to replacing hazardous X-ray systems with low-field PMRI scanners.
\end{abstract}

}

\maketitle

\IEEEdisplaynontitleabstractindextext

\section{Introduction}

\IEEEPARstart{L}{ow-Field} Magnetic Resonance Imaging (LF-MRI) is gaining momentum as an affordable alternative to clinical MRI, the current gold standard in numerous medical imaging applications, but also extremely expensive and often inaccessible \cite{Sarracanie2015,Marques2019,Sarracanie2020}. The main cost driver in an MRI scanner is the superconducting magnet required to generate the strong, static magnetic field ($B_0$) that enables the high quality images typical for clinical MRI. By lowering the field strength, the need for superconducting magnets is removed, resulting in a drastic reduction of the economic and energetic needs. On the other hand, the signal-to-noise ratio (SNR) of the magnetic resonance signals and reconstructed images is also greatly compromised.

Prepolarization is a long-established technique designed to partially compensate for the SNR loss in LF-MRI \cite{Macovski1993,Morgan1996,Kegler2007,Lee2005,Obungoloch2018}, and could be of special relevance for hard biological tissues where hydrogen content is sparse and signals decay very fast \cite{Algarin2020,Rigla2021}. In Prepolarized MRI (PMRI), the Boltzmann equilibrium magnetization of the sample is boosted by an intense, not necessarily homogeneous, magnetic pulse of amplitude $B_\text{p}$ before the start of the imaging pulse sequence, which is then executed at a lower but highly homogeneous $B_0$. For efficient PMRI, the prepolarization pulse must be turned off in a time $t_\text{off}$ much shorter than the sample $T_1$ relaxation time over which the extra magnetization is lost. This is easily met for liquids and soft biological tissues, where spin-lattice interactions are averaged out by the molecular tumbling of water, leading to relaxation times above 100\,ms \cite{BkKowalewski}. Indeed, PMRI has already demonstrated its potential for \emph{ex vivo} and \emph{in vivo} imaging of soft samples at field strengths ranging from hundreds of milli-tesla to hundreds of micro-tesla \cite{Rigla2021,Matter2006,Matter2006-2,Venook2006,Savukov2013,Inglis2013}. For solid-state matter or hard biological tissues (e.g. dental tissues), which feature short $T_1$ times, prepolarization is much more challenging: the suppressed proton mobility prevents the averaging-out of dipolar interactions by molecular tumbling of protons in water. This effect is even more pronounced at low field strengths, where the Larmor frequency is closer to proton tumbling frequencies \cite{BkDuer}. On the other hand, hard tissue PMRI could be of relevance for dental clinical practice, where hazardous X-ray systems are massively used \cite{Shah2014}, and for which there is no affordable MRI alternative as of yet \cite{Algarin2020,Mastrogiacomo2019,Idiyatullin2011,Weiger2012}. 

In this paper, we demonstrate prepolarization and imaging of samples with ultra-short $T_1$, down to a few tens of milli-seconds. After brief introductions to the relevant theoretical framework and experimental equipment in Secs.\,\ref{sec:theory} and \ref{sec:apparatus} respectively, we analyze in Sec.\,\ref{sec:sweeps} the signal strength boost for an inorganic solid-state sample as a function of pulse sequence parameters. Besides revealing the effect of prepolarization, this study also shows that the simple model presented in Sec.\,\ref{sec:theory} adequately describes the observed data, where deviations can be attributed to the effect of sudden mechanical displacements due to the strong interaction between the main and prepolarization magnets during the prepolarization pulse. In Sec.\,\ref{sec:teeth}, we present the first prepolarized magnetic resonance images (of a cattle bone and a human tooth), which show an SNR increase of a factor of 2 with respect to an equivalent acquisition without prepolarization. Finally, in Sec.\,\ref{sec:conc}, we discuss the feasibility of extending the presented MRI concept to clinical applications in the field of dentistry and orthodontics.


\section{Theory}
\label{sec:theory}

\begin{figure}
	\centering
	\includegraphics[width=1.\columnwidth]{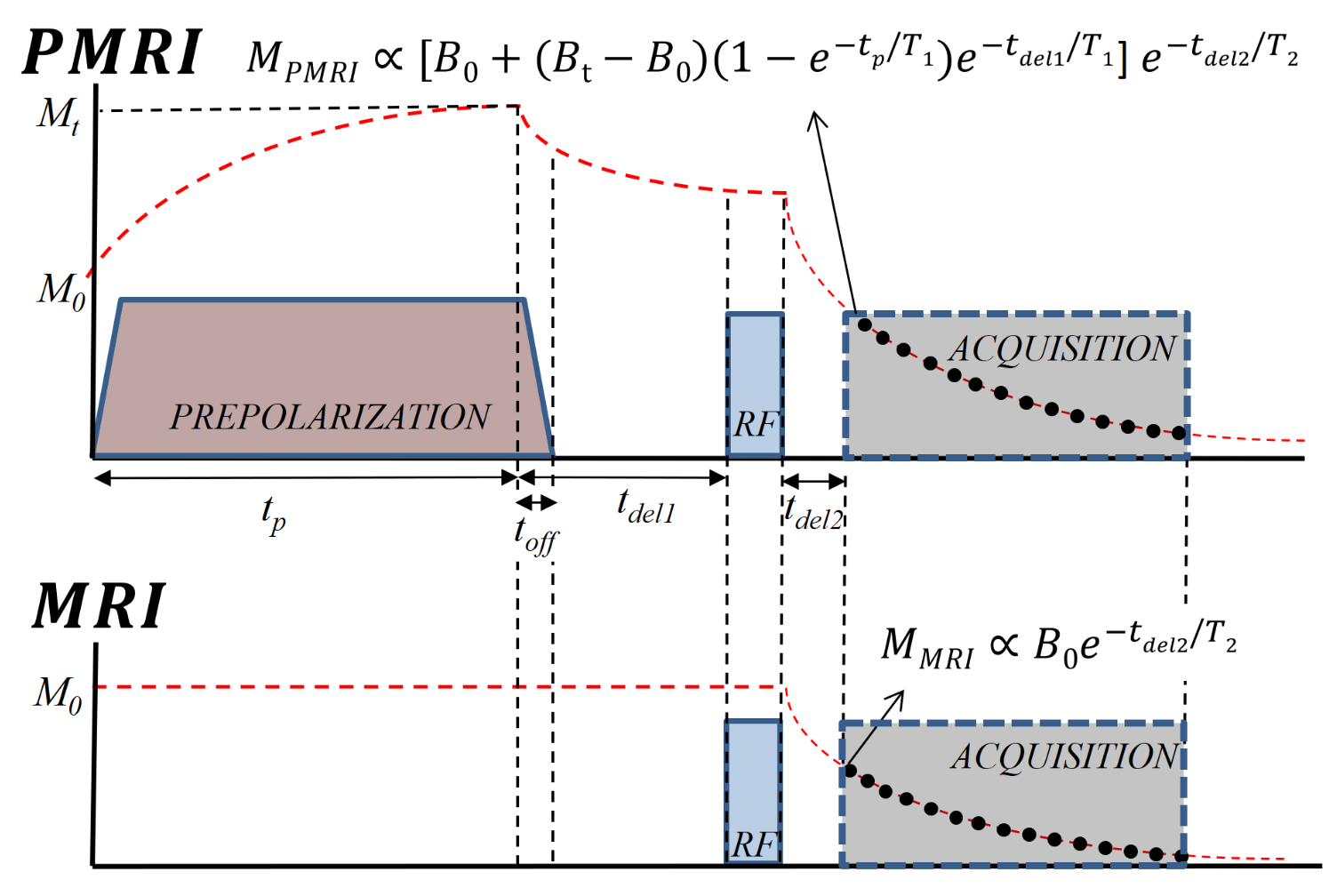}
	\caption{PMRI (top) and MRI (bottom) pulse sequences used in this work, with analytical expressions for the magnetization at the start of FID acquisition. Their ratio $\alpha$ represents the SNR gain due to prepolarization, as per Eq.\,(\ref{eq:alpha}). $M_\text{t}$ and $M_0$  are the magnetizations in thermal equilibrium with and without prepolarization and are directly proportional to $B_\text{t}$ and $B_0$ respectively.}
	\label{fig:seq}
\end{figure}

\begin{figure}
	\centering
	\includegraphics[width=1.\columnwidth]{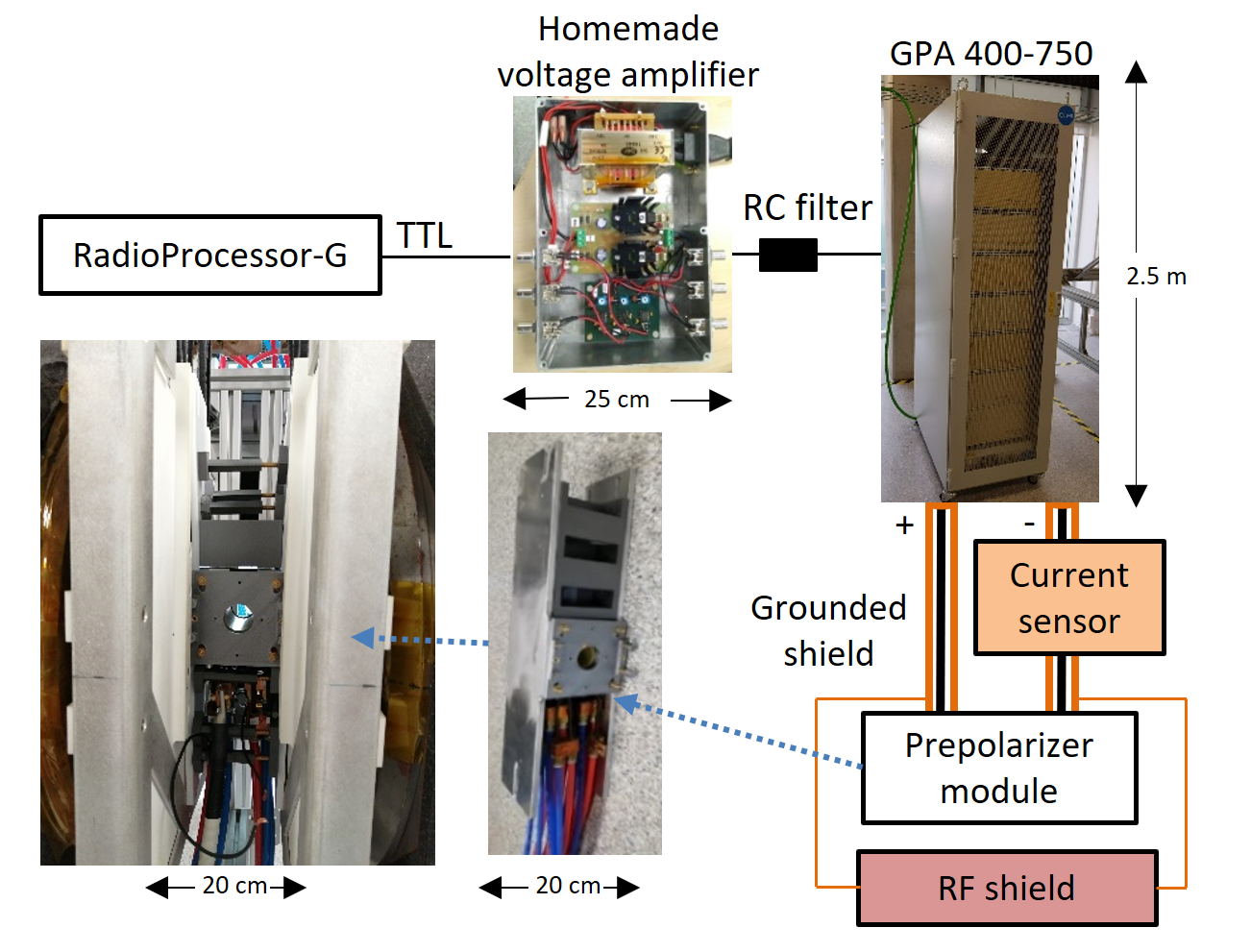}
	\caption{Scheme of the setup employed for hard tissue PMRI.}
	\label{fig:setup}
\end{figure}

To quantify the effect of the prepolarization on hard tissues, in the remainder of the paper we compare the signals resulting from magnetic pulse sequences based on those in Fig.\,\ref{fig:seq}. These sequences are identical except for the fact that the prepolarization pulse has an amplitude $B_\text{p}$ in the PMRI sequence and zero in the standard MRI sequence. For an homogeneous sample of characteristic relaxation time $T_1$, we define the prepolarization gain $\alpha$ as the ratio between the sample magnetizations during the data acquisitions:
\begin{eqnarray}
	\nonumber M_\text{PMRI} &\propto& \left( B_0 + (B_\text{t}-B_0) \left(1-\Exp{-t_\text{p}/T_1}\right) \Exp{-t_\text{del1}/T_1} \right) \Exp{-t_\text{del2}/T_2},\\
	 M_\text{MRI} &\propto& B_0 \Exp{-t_\text{del2}/T_2},
\end{eqnarray}
so
\begin{equation}\label{eq:alpha}
	\alpha \equiv \frac{M_\text{PMRI}}{M_\text{MRI}} = 1+\frac{B_\text{t}-B_0}{B_0} \left( 1-\Exp{-t_\text{p}/T_1} \right) \Exp{-t_\text{del1}/T_1},
\end{equation}
where we neglect the duration of RF pulses. Here: $B_\text{t} = |\vec{B}_0+\vec{B}_\text{p}|$ is the total field strength during the prepolarization pulse, where the main and prepolarization fields need not be parallel; $t_\text{p}$ is the prepolarization pulse length, during which the magnetization asymptotically reaches equilibrium with $B_\text{t}$; $t_\text{off}$ is the ramp down time of the prepolarization pulse; $t_\text{del1} \geq t_\text{off}$ is the time from the moment the prepolarization pulse starts to be switched off until the beginning of the radio-frequency (RF) excitation; $t_\text{del2}$ is the time between the RF pulse and the start of the data acquisition; and $T_2$ is the sample-dependent dephasing characteristic time over which the magnetization decoheres. Admittedly, this definition of SNR enhancement tends to overestimate the benefits of PMRI, since the standard MRI sequence could be shortened and its SNR increased by further averaging in the same overall acquisition time. Nevertheless, this is the simplest possible comparison and is typical in the literature (see e.g. \cite{Matter2006}).


\section{Apparatus}
\label{sec:apparatus}

As a result of the short $T_1$ timescales typical of solids, hard tissue prepolarization poses a significant engineering challenge to achieve fast enough $t_\text{off}$ times. Our solution to this follows.

The ``DentMRI - Gen I'' 0.26 T scanner and prepolarization modules employed for this work (see Fig.\,\ref{fig:setup}) are described in detail elsewhere \cite{Algarin2020,Rigla2021}. Essentially, our group has designed, built and characterized a prepolarizer coil whose main parameters of inductance, resistance and efficiency are $L\approx\SI{600}{\micro H}$, $R\approx\SI{75}{m \ohm}$ and $\eta\approx\SI{1.9}{mT/A}$. The gap between the planar gradient stacks is $\approx\SI{210}{mm}$, placing a hard boundary on the prepolarizer module size and, consequently, to the maximum achievable coil inductance. Due to geometric limitations and to ease accessibility, we placed the prepolarizer module so that $\vec{B}_\text{p}$ is perpendicular to $\vec{B}_0$ \cite{Rigla2021}. This reduces the maximum achievable $B_\text{t}$ from $|\vec{B}_0|+|\vec{B}_\text{p}|$ to $(B_0^2+B_\text{p}^2)^{1/2}$, but has the advantage that the generated Eddy currents and the residual energy in the prepolarization coil barely disturb the longitudinal field $\vec{B}_0$ (e.g. when $B_\text{p}$ falls to $1$\,mT, the total field deviates from the original $B_0$ by only \SI{2}{\micro T}).

\begin{figure}
	\centering
	\includegraphics[width=1.\columnwidth]{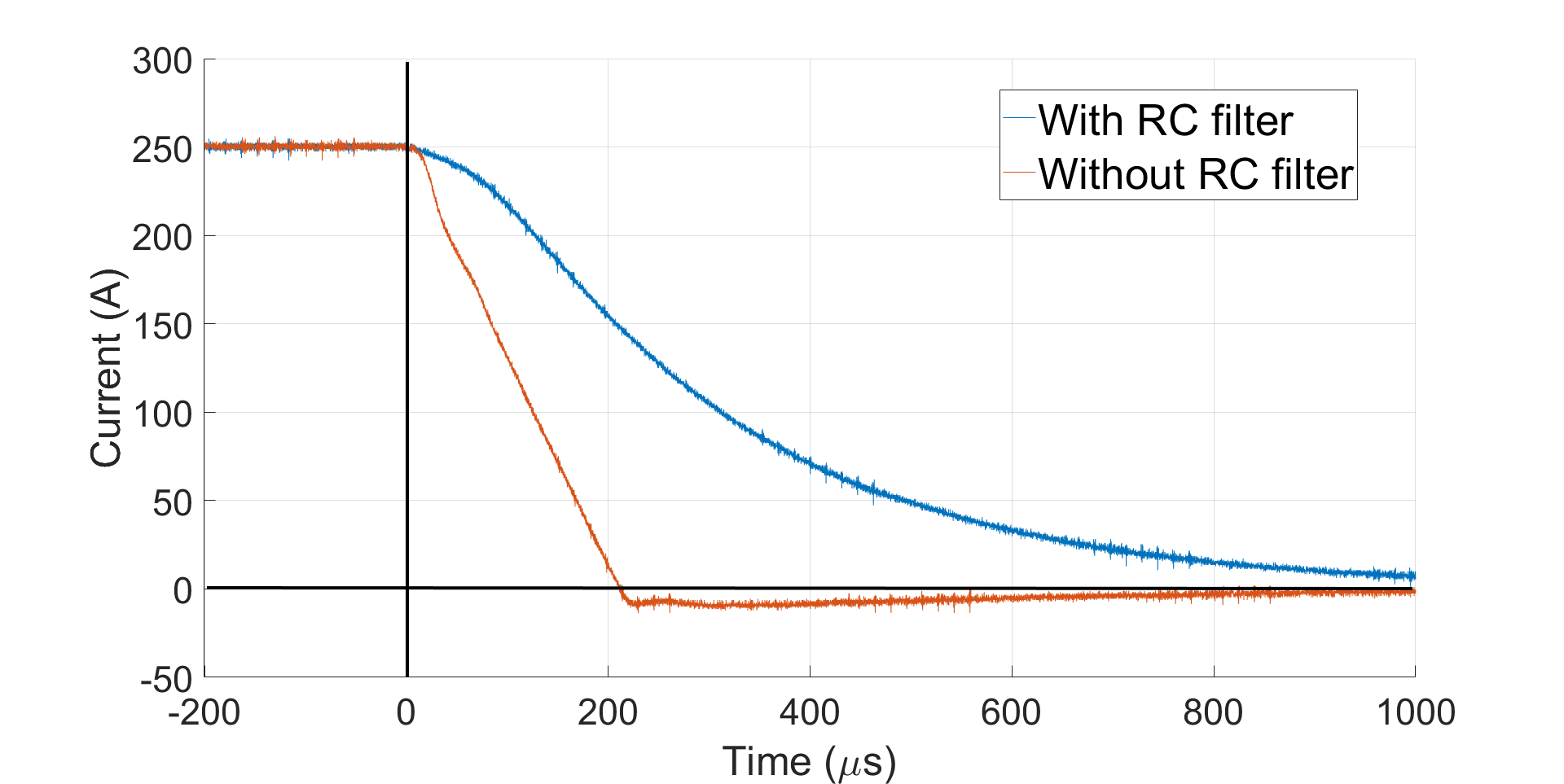}
	\caption{Falling edge of the prepolarization pulse current from 250\,A ($B_\text{p}\approx0.475$\,T) with the GPA 400-750, with and without the low-pass RC filter (see text). With the RC filter, the prepolarizer field is 1\,mT after $\approx 1.73$\,ms (not shown), so the deviation of the total field with respect to 0.26\,T is just \SI{2}{\micro T} and the Larmor frequency may be considered stabilized.}
	\label{fig:currenttransitions}
\end{figure}

In order to cope with the short $T_1$ of hard biological tissues, the high power electronics setup for the prepolarizer module has been substantially upgraded with respect to the system introduced in Ref.\,\cite{Rigla2021}. In the current apparatus, a digital output from the RadioProcessor-G board (SpinCore Electronics LLC) is amplified in two stages, first in a home-made variable-gain low-voltage amplifier, and then in a high power (\SI{400}{A} and \SI{750}{V}) gradient amplifier from International Electric Co. (GPA 400-750). The latter can ramp currents from 0 to $\pm$260\,A in $\approx\SI{200}{\micro s}$ in our $\approx\SI{600}{\micro H}$ load (see Fig. \ref{fig:currenttransitions}), where we were previously limited to $\approx\SI{35}{ms}$ \cite{Rigla2021}. Figure \ref{fig:currenttransitions} also shows a smoother transition corresponding to the case where we low-pass filter the digital output with an RC circuit of characteristic time constant $\approx\SI{350}{\micro s}$. We find this convenient to avoid mechanical stress in the module due to the sudden appearance of strong magnetic interactions between the main magnet and the prepolarizer. This reduces the generation of Eddy currents and, thereby, distortions in the acquired signals and image reconstructions due to uncontrolled magnetic field dynamics. All the measurements below are with the low-pass filter.

\section{SNR enhancement}
\label{sec:sweeps}


For calibration and first tests we employed a sample made of a photopolymer resin \cite{Rai2017}, which is highly homogeneous, abundant in hydrogen and features relaxation parameters comparable to the enamel in human teeth. At our $B_0$, we have measured $T_1\approx 23.1$\,ms and $T_2 \approx \SI{650}{\micro s}$ with Inversion Recovery \cite{Bydder1998} and CPMG \cite{Carr1954,Meiboom1958} pulse sequences, respectively.

\begin{figure}
	\centering
	\includegraphics[width=1.\columnwidth]{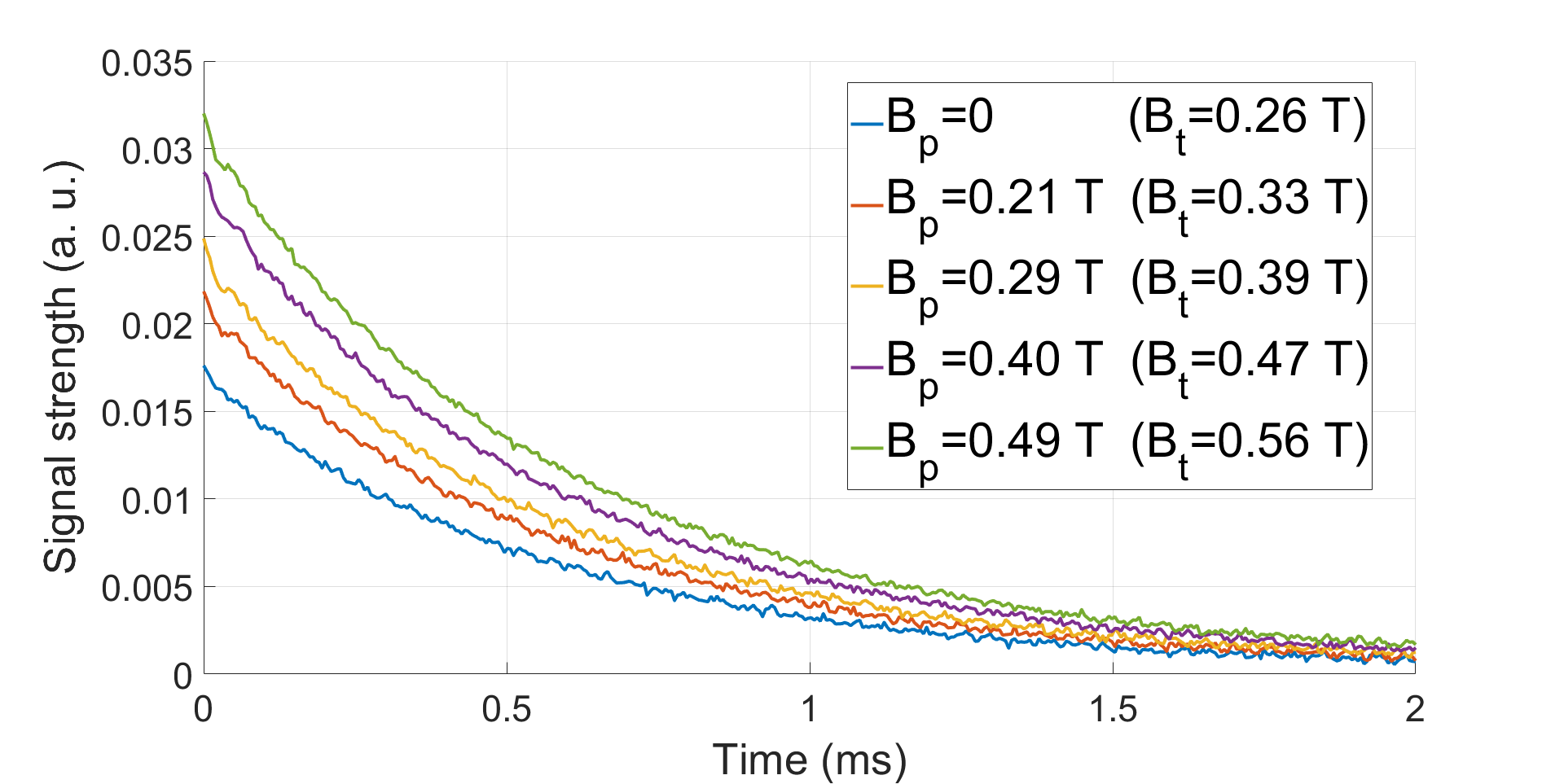}
	\caption{FIDs after prepolarizing the photopolymer resin sample with pulses of $t_\text{p}=160$\,ms and $t_\text{del1}=\SI{3}{ms}$ for different $B_\text{p}$ values.}
	\label{fig:test1}
\end{figure}

First we check whether the SNR is enhanced by prepolarization as predicted by the model in Eq.\,(\ref{eq:alpha}). To that end, we set $t_\text{p}=160$\,ms ($>7T_1$) in the sequence in Fig.\,\ref{fig:PulseSequence} to prepolarize close to the saturation magnetization. Next, a resonant $\pi/2$ RF pulse coherently rotates the magnetization to the transverse plane. Both pulses are separated by a wait time $t_\text{del1}=\SI{3}{ms}$, long enough to avoid Larmor frequency shifts and distortions in the acquired Free Induction Decay (FID) signals due to residual magnetic energy in the prepolarizer. The signal readout starts $t_\text{del2}=\SI{100}{\micro s}$ after the RF pulse to avoid ring-down from the RF coil. The resulting FID is acquired for $t_\text{acq}=2$\,ms with a readout bandwidth $BW=200$\,kHz. This protocol is repeated for four different voltage gains of our home-made amplifier, generating $B_\text{p} \approx0.21$, 0.29, 0.40 and 0.49\,T, which correspond to $B_\text{t} \approx0.33$, 0.39, 0.47 and 0.56\,T. Figure\,\ref{fig:test1} shows the absolute value of the FIDs for these cases and for the standard MRI sequence ($B_\text{p} = 0$ and $B_\text{t} \approx0.26$\,T). For a given value of $B_\text{p}$, we calculate the prepolarization boost $\bar{\alpha}_{B_\text{p}}$ as the mean ratio of the PMRI and standard MRI data:
\begin{equation}\label{eq:alphabar}
	\bar{\alpha}_{B_\text{p}} = \frac{1}{N_\text{points}}\sum_{i=1}^{N_\text{points}} \frac{s_{B_\text{p}}(t_i)}{s_0(t_i)},
\end{equation}
where $N_\text{points}=t_\text{acq}\cdot BW$, $s_{B_\text{p}}(t_i)$ is the signal amplitude measured for the PMRI with prepolarization strength $B_\text{p}$ for the time bin $(t_i)$, and $s_0(t_i)$ is the amplitude measured for the standard MRI sequence at $t_i$. The estimated $\bar{\alpha}_{B_\text{p}}$ values are $1.240\pm 0.005$, $1.430\pm 0.008$, $1.705\pm 0.008$ and $1.964\pm 0.011$ for the above prepolarization field strengths, where the given uncertainties indicate the standard error of the mean
\begin{equation}\label{eq:sigma}
	\sigma_{\bar{\alpha}} = \frac{1}{N_\text{points}}\sqrt{\sum_{i=1}^{N_\text{points}} \left( \frac{s_{B_\text{p}}(t_i)}{s_0(t_i)} - \bar{\alpha}_{B_\text{p}} \right)^2}.
\end{equation}
The corresponding theoretical $\alpha$ values for $T_1\approx 23.1$\,ms can be calculated from Eq.\,(\ref{eq:alpha}): $\alpha\approx$ 1.24, 1.44, 1.72 and 1.98.

\begin{figure*}
	\centering
	\includegraphics[width=2.\columnwidth]{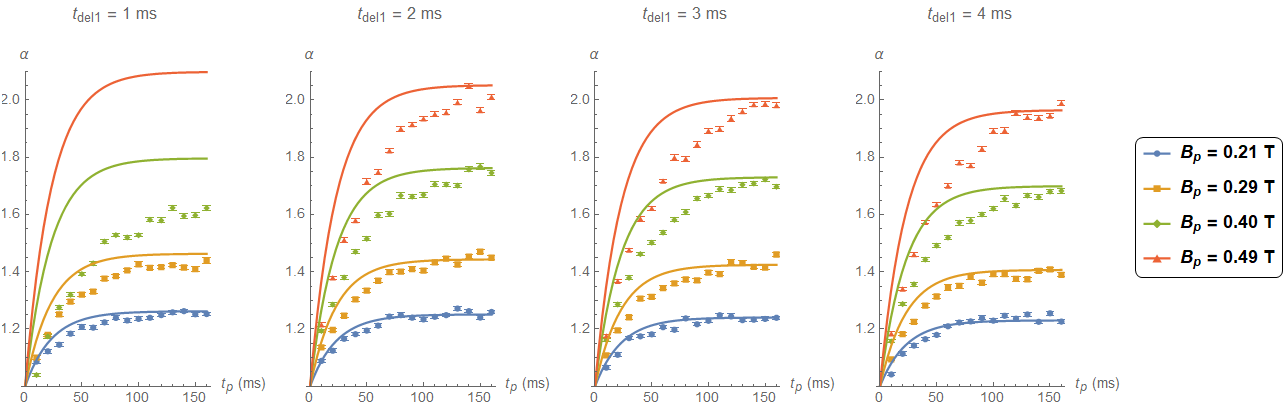}
	\caption{Comparison between theoretical (continuous lines) and experimental (data points) gain $\alpha$ for different values of $t_\text{p}$, $B_\text{p}$ and $t_\text{del1}$, using the photopolymer resin sample. The data for $B_\text{p}=0.49$\,T is not included because it was heavily corrupted by the sharp magnetic transitions (see text). For a given $B_\text{p}$, the maximum gain decreases with $t_\text{del1}$ due to $T_1$ decay after the prepolarization pulse.}
	\label{fig:test2}
\end{figure*}

The small experimental deviations from the theoretically calculated values could arise from: i) mechanical vibrations due to magnetic forces, ii) induced Eddy currents or iii) off-resonant spin evolution due to a time-dependent Larmor frequency. All three are more pronounced for intense $B_\text{p}$ values and short $t_\text{del1}$ times. To find a working regime free of these effects, we have characterized their influence on the SNR gain with the measurements shown in Fig.\,\ref{fig:test2}.

For the plots in Fig.\,\ref{fig:test2} we sweep the prepolarization pulse duration from $t_\text{p}=10$ to 160\,ms and $t_\text{del1}$ from 1 to 4\,ms, for the same four $B_\text{p}$ values as above. The gain and uncertainty for every data point are estimated according to Eqs.\,(\ref{eq:alphabar}) and (\ref{eq:sigma}). The solid lines in the figure correspond to calculations employing the model in Eq.\,(\ref{eq:alpha}).

Unsurprisingly, for the weaker prepolarization currents we measure FID curves that follow closely theoretical predictions, even for $t_\text{del1}$ as short as 1\,ms. Deviations are stronger for short wait and prepolarization times. In the extreme case of $B_\text{p}\approx0.49$\,T and $t_\text{del1}=1$\,ms, the measured data was heavily corrupted and did not follow the typical exponential behavior (i.e. as in the FIDs in Fig.\,\ref{fig:test1}). It is unlikely that these issues are due to drifts in the Larmor frequency as the prepolarizer relaxes, since a residual orthogonal field perturbs $B_0$ very weakly (e.g., for $B_\text{p}\approx 0.49$\,T and $t_\text{del1}=1$\,ms, the Larmor frequency shifts by only 250 Hz). On the other hand, Eddy currents and especially mechanical vibrations can be behind for the aforementioned deviations. In fact, we have observed that these unwanted effects are more prominent if the prepolarizer is not rigidly fixed to the scanner. With the mechanical fixation in place (see Fig.\,\ref{fig:setup}), the system performs well away from this extreme regime. Indeed, the plots in Fig.\,\ref{fig:test2} demonstrate that the measured SNR gain is compatible with theoretical predictions for prepolarization pulses longer than 120\,ms and $t_\text{del1}\geq2$\,ms.


\section{Hard tissue PMRI}
\label{sec:teeth}
\begin{figure*}
	\centering
	\includegraphics[width=2.\columnwidth]{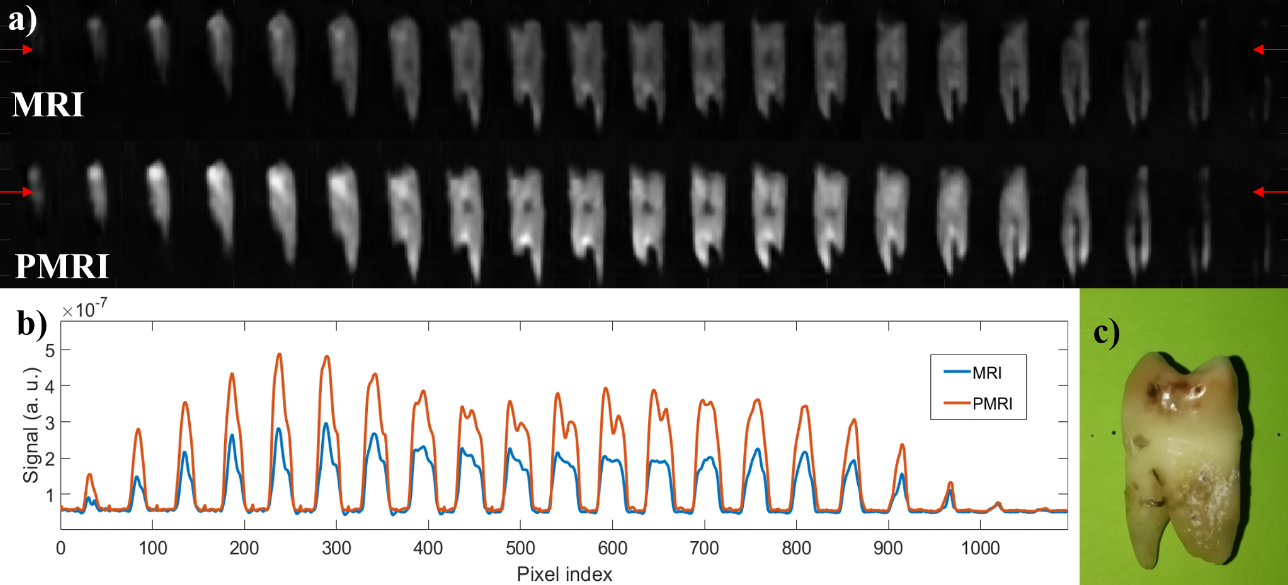}
	\caption{(a) PETRA (top) and P-PETRA (bottom) images of an \emph{ex-vivo} adult human molar tooth. (b) Signal intensity along the horizontal line defined by the red arrows in (a). The experimentally obtained value for the prepolarization gain is $\alpha\approx1.97$ (expected value $\approx2.02$, see text). (c) Photograph of the sample.}
	\label{fig:dientesPMRI}
\end{figure*}

\begin{figure*}
	\centering
	\includegraphics[width=2.\columnwidth]{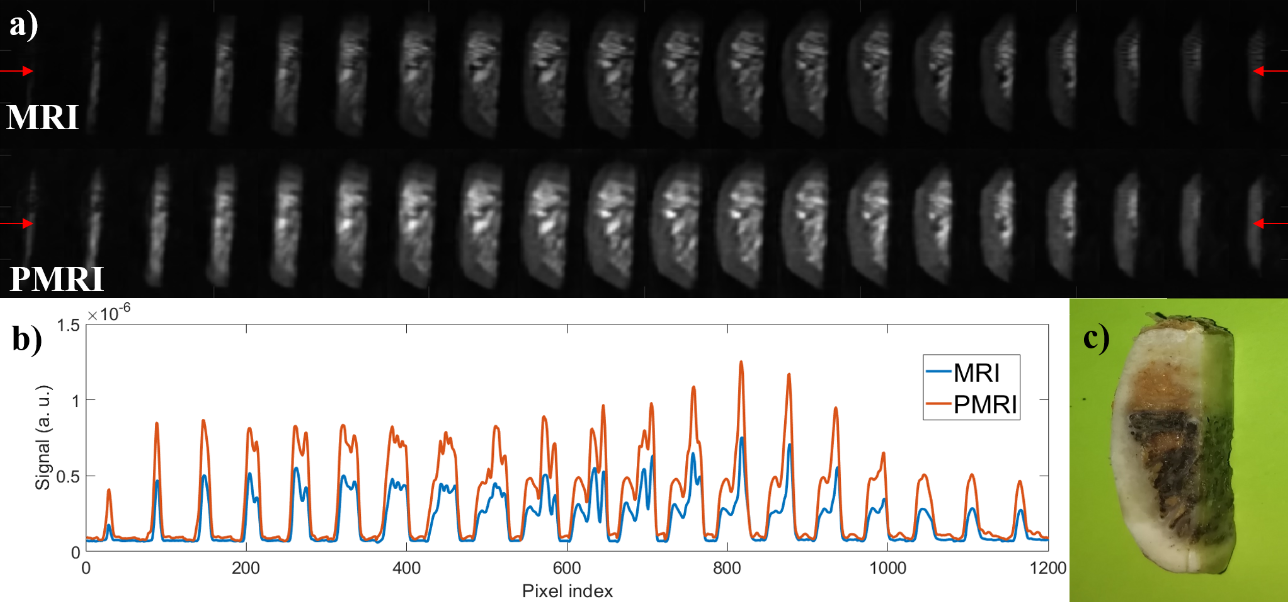}
	\caption{(a) PETRA (top) and P-PETRA (bottom) images of an \emph{ex-vivo} piece of cattle rib bone. (b) Signal intensity along the horizontal line defined by the red arrows in (a). The experimentally obtained value for the prepolarization SNR gain is $\alpha\approx1.99$ (expected value $\approx2.00$, see text). (c) Photograph of the sample.}
	\label{fig:huesoPMRI}
\end{figure*}

In this section we demonstrate the system's capability for imaging hard biological tissues with PMRI. To that end, we employ: i) an adult human molar tooth (Fig. \ref{fig:dientesPMRI}(c)) extracted one year before these experiments and dried so that primarily mineralized matter (dentin and enamel) remains; and ii) a piece of cattle rib (Fig. \ref{fig:huesoPMRI}(c)) including cortical and spongy bone tissues. We have measured the $T_1$ times of both samples by Inversion Recovery, and found $T_1 \approx20.3$ and \SI{19.3}{ms} for the tooth and bone, respectively. The cattle bone contains both cortical and spongy tissues, so the estimated time is an averaged quantity. The $T_1$ times of all the employed samples are very similar, so we can determine suitable parameter regimes from the measurements on the photopolymer resin (Fig.\,\ref{fig:test2}).

\begin{figure}
	\centering
	\includegraphics[width=1.\columnwidth]{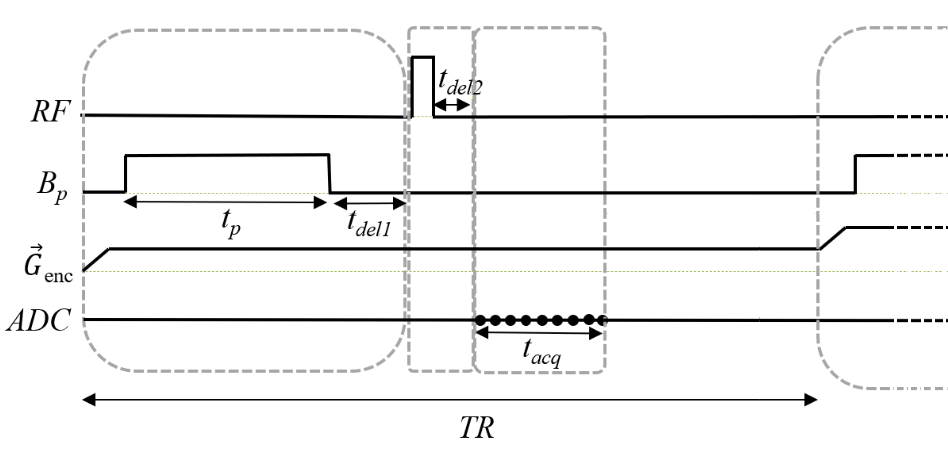}
	\caption{P-PETRA pulse sequence integrating the PMRI sequence in Fig.\,\ref{fig:seq} with PETRA. P-PETRA is employed for the prepolarized hard tissue images in Figs.\,\ref{fig:dientesPMRI} and \ref{fig:huesoPMRI}. Here, $\vec{G}_\text{enc}$ is the frequency encoding gradient and the ADC (analog-to-digital converter) acquisition is marked with black points.}
	\label{fig:PulseSequence}
\end{figure}

The ultra-short $T_2$ times typical of hard tissues impose the use of dedicated MRI sequences, such as those in the Zero Echo Time (ZTE) family \cite{Weiger2013}. These are characterized by radial $k$-space acquisitions beginning immediately after the RF excitation, to capture as much as possible of the short-lived signal. Ramping the gradient is time consuming, so in ZTE sequences the spatial encoding gradients are switched on before the RF pulse. In this work, we even switch on the frequency encoding gradient before prepolarization \cite{Kobayashi2015} to limit mechanical vibrations and the influence of Eddy currents during acquisition. Having the gradient on during resonant excitation imposes the use of hard (short and intense) RF pulses, leading to spurious signals which could corrupt the data acquisition. To prevent this, we introduce a delay $t_\text{del2}$ before the readout, resulting in a gap without data at the center of $k$-space. This can be filled with additional acquisitions \cite{Weiger2019}. One possibility is to do so is in a pointwise fashion, as in PETRA (Pointwise Encoding Time-reduction with Radial Acquisition, \cite{Grodzki2012}). For the following images we employ a PETRA sequence with a prepolarization stage before the RF excitation (P-PETRA, Fig. \ref{fig:PulseSequence}).

In Fig.\,\ref{fig:dientesPMRI} we show prepolarized images of a human molar tooth  obtained following the scheme  in Fig.\,\ref{fig:PulseSequence}. The size of the field of view is set to $21\times13\times\SI{13}{mm^3}$ and the image is reconstructed with Algebraic Reconstruction Techniques (ART, \cite{Algarin2020,Karczmarz1937,Gower2015}) into $42\times26\times26$ voxels. The acquisition starts $t_\text{del1}=\SI{130}{\micro s}$ after the RF pulse to avoid the effect of ring-down and lasts $t_\text{acq}=\SI{700}{\micro s}$, with a bandwidth $BW\approx\SI{30}{kHz}$. The repetition time is set to $\text{TR}=\SI{250}{ms}$, limited by the maximum duty cycle of the GPA 400-750 at this current regime. We undersample the number of radial lines in $k$-space by a factor $\times 8$ with respect to the Nyquist criterion, where ART reconstructions are still robust. Every image contains 12 averages for a total scan time of $\approx\SI{29}{min}$. The bottom row of images in Fig.\,\ref{fig:dientesPMRI}(a) corresponds to scans in which a prepolarization pulse is triggered with a current intensity of $\approx$260\,A ($B_\text{t} \approx$ 0.56\,T), which lasts $t_\text{p}=90$\,ms and where $t_\text{del1}=\SI{2}{ms}$. The pulse sequence for the top row of Fig.\,\ref{fig:dientesPMRI}(a) is identical, but the prepolarization pulse is not triggered ($B_\text{p}=0$, $B_\text{t}=0.26$\,T). The brightness scale is common to both datasets to highlight the gain in SNR with PMRI. Both images have been denoised using a Block-Matching filter \cite{Algarin2020,Maggioni2013}. To quantify the influence of prepolarization, we plot in Fig.\,\ref{fig:dientesPMRI}(b) the same profile along a horizontal line around the upper portion of the images in (a), in the region of the tooth crown. The mean $\alpha=\text{SNR}_\text{PMRI}/\text{SNR}_\text{MRI}$ (before filtering and averaged over a region of interest of constant bright pixels around the dentin) is $\approx1.97$, where $\text{SNR}_\text{PMRI}=\bar{s}_\text{PMRI}/\bar{n}_\text{PMRI}\approx16.46$, and $\text{SNR}_\text{MRI}$ (analogously defined) is $\approx8.36$. The mean signal and noise values ($\bar{s}$ and $\bar{n}$) are estimated, respectively, as the mean value and standard deviation of the voxel brightness in the region of interest. For comparison, the expected prepolarization gain from Eq.\,(\ref{eq:alpha}) is $\approx2.02$.

We have applied an analogous protocol to image a piece of a cattle rib bone. The size of the field of view is set to $36\times15\times\SI{15}{mm^3}$ and the image is reconstructed with ART into $72\times30\times30$ voxels. The acquisition starts $t_\text{del2}=\SI{125}{\micro s}$ after the RF pulse and lasts $t_\text{acq}=\SI{800}{\micro s}$, with a bandwidth $BW\approx\SI{45}{kHz}$. The repetition time is $\text{TR}=\SI{280}{ms}$. The $k$-space undersampling is again $\times 8$. Every image contains 11 averages for a total scan time of $\approx\SI{53}{min}$. The bottom row of images in Fig.\,\ref{fig:huesoPMRI}(a) corresponds to scans in which a prepolarization pulse is triggered with a current intensity of $\approx260$\,A ($B_\text{t} \approx0.56$\,T), which lasts $t_\text{p}=90$\,ms and where $t_\text{del1}=\SI{1.5}{ms}$. The pulse sequence for the top row of Fig.\,\ref{fig:huesoPMRI}(a) is identical, but the prepolarization pulse is not triggered ($B_\text{p}=0$, $B_\text{t}=0.26$\,T). The brightness scale is again common to both datasets, and the images have been also Block-Matched filtered. The SNR enhancement is evident in Fig.\,\ref{fig:dientesPMRI}(b), which shows the reconstructed signal intensity profile along a horizontal line around the middle region of the images in (a). The measured mean $\alpha=\text{SNR}_\text{PMRI}/\text{SNR}_\text{MRI}$ is $\approx1.99$, where $\text{SNR}_\text{PMRI} \approx35.5$ and $\text{SNR}_\text{MRI}$ $\approx17.8$ (defined as in the previous paragraph). For comparison, the expected prepolarization gain from Eq.\,(\ref{eq:alpha}) is $\approx2.00$.


\section{Conclusion and outlook}
\label{sec:conc}
We have shown that it is possible to enhance the  quality of magnetic resonance images of hard tissues at low magnetic fields by means of a high power prepolarizer module, for a total cost of $\approx20$\,k\euro, where the GPA 400-750 module is around 13\,k\euro. The major challenges we have faced are: i)  integrating a high power drive capable of switching off the prepolarization pulse fast enough; and ii) coping with mechanical vibrations due to the strong magnetic interaction between the main and prepolarization fields.

The preliminary results shown in this work have been obtained in a highly constrained setup in terms of prepolarizer alignment, hydraulic capacity and prepolarizer duty cycle. If the prepolarization field were aligned with the main static field, we could have approached $B_\text{t}=0.74$\,T, leading to an increase in SNR of $\times 2.85$. Also, limitations in the cooling system forced us to work under 260\,A, where the system could have taken up to 320\,A. This corresponds to $B_\text{t}\approx0.66$\,T with the current configuration, or $B_\text{t}$ $\approx0.92$\,T if $\vec{B}_0$ and $\vec{B}_\text{p}$ are aligned. A further limitation of our setup is the maximum duty cycle of the GPA 400-750 module, which enforces repetition times $\text{TR}\geq\SI{250}{ms}$. These are significantly longer than strictly required by the $T_1$ values of the samples. Assuming a hypothetical $\text{TR}\geq 4  T_1$, enough to thermalize at 98\,\% of the longitudinal magnetization, $\text{TR}=60$\,ms would have sufficed for prepolarization of teeth. Without these limitations, i.e. with $\text{TR}=60$\,ms (shorter acquisitions), $I_\text{p}=320$\,A and $\vec{B}_\text{p}||\vec{B}_\text{0}$, we could achieve $B_\text{t}\approx 0.92$\,T and $\alpha\approx3.5$, compared to $B_\text{t}\approx 0.56$\,T and $\alpha\approx2$.

The results in this paper are of potential application to clinical dental MRI. This would require a prepolarizer magnet large enough to fit a human head. Matter \emph{et al.} made a 0.4\,T prepolarizer of $\approx127$\,mm in diameter, which they used for \emph{in vivo} PMRI of a human wrist \cite{Matter2006}. We argue next that a larger coil for dental applications is also realistic. The magnetic field strength inside a solenoid of inner (outer) radius $r_\text{in}$ ($r_\text{out}$) and length $l$ is given by
\begin{equation}
	B_\text{p}=\mu_0 G(\alpha,\beta) \sqrt{\frac{P \lambda}{\rho r_\text{in}}},
\end{equation}
where $\mu_0$ is the vacuum permeability, $P$ is the power dissipated in the coil due to resistive losses, $\lambda$ is the fraction of conductor material in the solenoid (to account for water refrigeration conducts, isolating material and gaps between windings and layers), $\rho$ is the resistivity of the conductor, and $G(\alpha,\beta)$ is a geometric factor defined as
\begin{equation}
	G(\alpha,\beta) = \sqrt{\frac{\beta}{2\pi(\alpha^2-1)}} \left( \sinh^{-1}(\alpha/\beta) - \sinh^{-1}(1/\beta) \right),
\end{equation}
with $\alpha = r_\text{out}/r_\text{in}$ and $\beta = l/(2r_\text{in})$ \cite{Matter2006,Montgomery1969}. Assuming the same copper wire as in Ref.\,\cite{Matter2006} (square section of side 4\,mm with a hole of radius 1\,mm), a solenoid with $n_\text{l}=7$ layers with $n_\text{w}=55$ windings each would have a total resistance $R\approx\SI{0.41}{\ohm}$ for $r_\text{in}=115$\,mm, $r_\text{out}=143$\,mm and $l=220$\,mm. For a drive current $I_\text{p} = 210$\,A, the dissipated power is $P=R I_\text{p}^2 \approx 18$\,kW and $B_\text{p} \approx 0.3$\,T. For comparison, the wrist coil in Ref.\,\cite{Matter2006} produces 0.4\,T at 16\,kW. The inductance of the prepolarizer coil can be estimated as \cite{ARRL1999}
\begin{equation}
	L_\text{p} \approx \SI{7.87}{\micro H}\times\frac{(r_\text{out} + r_\text{in})^2 n_\text{w}^2 n_\text{l}^2}{3(r_\text{out} + r_\text{in}) + 9l + 10(r_\text{out} - r_\text{in})},
\end{equation}
where all distances must be given in meters. Using the above numbers we find $L_\text{p}\approx26$\,mH. With the 750\,V available from the GPA 400-750 unit, the current could be switched off in a time $t_\text{off}\approx 7$\,ms, still significantly shorter than the $T_1$ of the hardest human tissues. At these field variation rates (50\,T/s), unwanted magneto-stimulation effects may take place \cite{IEC2010}. This can be further investigated in dedicated setups \cite{Grau-Ruiz2020} and, if required, the prepolarization coil could be designed specifically to avoid peripheral nerve stimulation \cite{Davids2017,Davids2019}.

\section*{Contributions}
The high power electronics for prepolarization were designed and installed by JMG, JB, JPR and JA. The prepolarizer and mechanical holder were designed, assembled and characterized by JPR, JMG, EP and JB, with contributions from DGR and JA. Experimental data in the ``DentMRI - Gen I'' scanner were taken by JB and JMG, with help from JMA, FG, RP and JA. Data analysis performed by JB and JMG, with input from JMA, FG, RP and JA. Animal handling and manipulation of biological tissues performed by JB. The paper was written by JB, FG and JA, with input from all authors. Experiments conceived by JMB, JA and AR.

\section*{Acknowledgment}
This work was supported by the Ministerio de Ciencia e Innovaci\'on of Spain through research grant PID2019-111436RB-C21. Action co-financed by the European Union through the Programa Operativo del Fondo Europeo de Desarrollo Regional (FEDER) of the Comunitat Valenciana 2014-2020 (IDIFEDER/2018/022). JMG and JB acknowledge support from the Innodocto program of the Agencia Valenciana de la Innovaci\'on (INNTA3/2020/22 and INNTA3/2021/17).

\section*{Ethical statement}
All animal parts were obtained from a local butcher and research was conducted following the 3R principles. Experiments using human teeth were approved by the medical center Cl\'inica Llobell Cortell S.L. Procedures were conducted following the approved protocols, and informed consent was obtained from participants prior to study commencement.

\bibliography{myrefs}

\end{document}